\documentstyle[aps,version2,preprint]{revtex}
\begin{document}
\draft
\preprint{OCIP/C 94-7}
\preprint{July 1994}
\begin{title}
Determination of Leptoquark Properties in Polarized $e\gamma$ Collisions
\end{title}
\author{Michael A. Doncheski and Stephen Godfrey}
\begin{instit}
Ottawa-Carleton Institute for Physics \\
Department of Physics, Carleton University, Ottawa CANADA, K1S 5B6
\end{instit}

\begin{abstract}
We study leptoquark production using polarized $e\gamma$ colliders for the
center of mass energies $\sqrt s=500$~GeV and 1~TeV.  We show that using
polarization asymmetries the ten different types of leptoquarks listed by
Buchm\"uller, R\"uckl and Wyler can be distinquished from one another for
leptoquark masses essentially up to the kinematic limit of the respective
colliders.  Thus, if a leptoquark were discovered an $e\gamma$ collider could
play a crucial role in determining its origins.
\end{abstract}

There is much interest in the study of leptoquarks (LQs), colour
(anti-)triplet, spin 0 or 1 particles, which carry both baryon and lepton
quantum numbers.  Such objects appear in a large number of extensions of the
standard model such as grand unified theories, technicolour, and composite
models.  Quite generally, the signature for leptoquarks is quite striking: a
high $p_{_T}$ lepton balanced by a jet (or missing $p_{_T}$ balanced by a jet,
for the $\nu q$ decay mode, if applicable).  Although the discovery of a
leptoquark would be dramatic evidence for physics beyond the standard model it
would lead to the question of which model the leptoquark originated from.
Given the large number of leptoquark types it would be imperative to measure
its properties to answer this question.

Following the notation of Buchm\"uller, R\"uckl and Wyler (BRW) \cite{buch},
the complete set of possible LQs numbers 10 is: $S_1$, $\tilde{S}_1$ (scalar,
iso-singlet); $R_2$, $\tilde{R}_2$ (scalar, iso-doublet); $S_3$ (scalar,
iso-triplet); $U_1$, $\tilde{U}_1$ (vector, iso-singlet); $V_2$, $\tilde{V}_2$
(vector, iso-doublet); $U_3$ (vector, iso-triplet).  The production and
corresponding decay signatures are quite similar, though not identical, and
have been studied separately by many authors.  Even focussing only on the Next
Linear Collider ($e^+ e^-$, $e \gamma$ and $\gamma \gamma$ modes), there is a
considerable number of works in the literature
\cite{leptoLN,HP,BLN,me,eboli,mad,aliev}.  The question arises as to how to
differentiate between the different types.  We propose to use a polarized
$e\gamma$ collider to differentiate the LQs: a polarized $e$ beam (like SLC)
in conjunction with a polarized-laser backscattered photon beam.  We
concentrate on LQ production in $e \gamma$, which makes use of the fact that
the hadronic component of the photon is important and cannot be neglected
\cite{me,photont,photone}.  Although the production of LQs in polarized
$e\gamma$ collisions was considered first in Ref.~\cite{aliev}, those authors
do not use the polarization information to determine the specific model of LQ.

We will assume that a peak in the $e + jet$ invariant mass is observed in some
collider ({\it i.e.}, the existence of {\bf a} LQ has been established), and
so we need simply to identify the particular type of LQ. We assume that the
leptoquark charge has not been determined and assume no intergenerational
couplings. Furthermore, we will assume that only one of the ten possible types
of LQs is present.  Table~2 of BRW gives information on the couplings to
various quark and lepton combinations; the missing (and necessary) bit of
information is that the quark and lepton have the same helicity (RR or LL) for
scalar LQ production while they have opposite helicity (RL or LR) for vector
LQ production.  It is then possible to construct the cross sections for the
various helicity combinations and consequently the double spin asymmetry, for
the different types of LQs.

We'll denote the various helicity cross sections as
$\sigma^{\lambda_e \lambda_q}$, $\lambda_i = +$ for R heliciy, $\lambda_i = -$
for L helicity and $\sigma_{TOT} = \sigma^{++} + \sigma^{+-} + \sigma^{-+} +
\sigma^{--}$.  As is usual in the case of polarized collider phenomenology, it
is useful to introduce the double longitudinal spin asymmetry $A_{LL}$:
\begin{equation}
A_{LL} = \frac{(\sigma^{++} + \sigma^{--}) - (\sigma^{+-} + \sigma^{-+})}
{(\sigma^{++} + \sigma^{--}) + (\sigma^{+-} + \sigma^{-+})}
\end{equation}
and the helicity sum and difference distribution functions of partons within
the photon:
\begin{eqnarray}
f_{q/\gamma}(x,Q^2) &=& f^+_{q/\gamma}(x,Q^2) + f^-_{q/\gamma}(x,Q^2)
\nonumber \\
\Delta f_{q/\gamma}(x,Q^2) &=& f^+_{q/\gamma}(x,Q^2) - f^-_{q/\gamma}(x,Q^2)
\end{eqnarray}
where $f^{+(-)}_{q/\gamma}(x,Q^2)$ is the probability of a quark with the same
(opposite) helicity as the photon to carry a fraction $x$ of the photon's
momentum.  It is then straightforward to construct $A_{LL}$ for all 10 types
of LQs in terms of $\Delta f_{q/\gamma}(x,Q^2) =
\Delta f_{\bar{q}/\gamma}(x,Q^2)$, $f_{q/\gamma}(x,Q^2) =
f_{\bar{q}/\gamma}(x,Q^2)$ and $\kappa_{L,R}$ ($g^2_{L,R}/4 \pi = \kappa_{L,R}
\alpha_{em}$).  There are 3 general cases:
\begin{itemize}
\item[1)] $\sigma^{+ \lambda_q} = 0$ (only left-handed electrons couple to LQ):
\begin{eqnarray}
A_{LL} (S_3) & = &
\frac{\int_{M_2/s}^1 dx/x
[ \Delta f_{u/\gamma}(y,M^2) + 2 \Delta f_{d/\gamma}(y,M^2) ]
f_{\gamma/e}(x)}
{\int_{M_2/s}^1 dx/x
[ f_{u/\gamma}(y,M^2) + 2 f_{d/\gamma}(y,M^2) ]
f_{\gamma/e}(x)} \\
A_{LL} (\tilde{V}_2) & = &
-\frac{\int_{M_2/s}^1 dx/x
[ \Delta f_{u/\gamma}(y,M^2)]
f_{\gamma/e}(x) }
{\int_{M_2/s}^1 dx/x
[ f_{u/\gamma}(y,M^2) ]
f_{\gamma/e}(x) } \\
A_{LL} (U_3) & = &
-\frac{\int_{M_2/s}^1 dx/x
[ \Delta f_{d/\gamma}(y,M^2) + 2 \Delta f_{u/\gamma}(y,M^2) ]
f_{\gamma/e}(x)}
{\int_{M_2/s}^1 dx/x
[ f_{d/\gamma}(y,M^2) + 2 f_{u/\gamma}(y,M^2) ]
f_{\gamma/e}(x)} \\
A_{LL} (\tilde{R}_2) & = &
\frac{\int_{M_2/s}^1 dx/x
[ \Delta f_{d/\gamma}(y,M^2) ]
f_{\gamma/e}(x) }
{\int_{M_2/s}^1 dx/x
[ f_{d/\gamma}(y,M^2) ]
f_{\gamma/e}(x)}
\end{eqnarray}
\item[2)] $\sigma^{- \lambda_q} = 0$ (only right-handed electrons couple to
LQ):
\begin{eqnarray}
A_{LL} (\tilde{S}_1) & = &
\frac{\int_{M_2/s}^1 dx/x
[\Delta f_{d/\gamma}(y,M^2)]
f_{\gamma/e}(x)}
{\int_{M_2/s}^1 dx/x
[ f_{d/\gamma}(y,M^2)]
f_{\gamma/e}(x)} \\
A_{LL} (\tilde{U}_1) & = &
-\frac{\int_{M_2/s}^1 dx/x
[ \Delta f_{u/\gamma}(y,M^2) ]
f_{\gamma/e}(x)}
{\int_{M_2/s}^1 dx/x
[ f_{u/\gamma}(y,M^2) ]
f_{\gamma/e}(x)}
\end{eqnarray}
\item[3)] $\sigma^{- \lambda_q},\sigma^{+ \lambda_\gamma} \neq 0$ (both right-
and left-handed electrons couple to LQ):
\begin{eqnarray}
A_{LL} (S_1) & = &
\frac{\int_{M_2/s}^1 dx/x
[ \Delta f_{u/\gamma}(y,M^2) ]
f_{\gamma/e}(x)}
{\int_{M_2/s}^1 dx/x
[ f_{u/\gamma}(y,M^2) ]
f_{\gamma/e}(x)} \\
A_{LL} (V_2) & = &
-\frac{\int_{M_2/s}^1 dx/x
[ \kappa_R(\Delta f_{u/\gamma}(y,M^2) +
\Delta f_{d/\gamma}(y,M^2)) +
\kappa_L \Delta f_{d/\gamma}(y,M^2) ]
f_{\gamma/e}(x)}
{\int_{M_2/s}^1 dx/x
[ \kappa_R(f_{u/\gamma}(y,M^2) + f_{d/\gamma}(y,M^2)) +
\kappa_L f_{d/\gamma}(y,M^2) ]
f_{\gamma/e}(x)} \\
A_{LL} (U_1) & = &
-\frac{\int_{M_2/s}^1 dx/x
[ \Delta f_{d/\gamma}(y,M^2) ]
f_{\gamma/e}(x)}
{\int_{M_2/s}^1 dx/x
[ f_{d/\gamma}(y,M^2) ]
f_{\gamma/e}(x)} \\
A_{LL} (R_2) & = &
\frac{\int_{M_2/s}^1 dx/x
[ \kappa_R(\Delta f_{u/\gamma}(y,M^2) +
\Delta f_{d/\gamma}(y,M^2)) +
\kappa_L \Delta f_{u/\gamma}(y,M^2) ]
f_{\gamma/e}(x)}
{\int_{M_2/s}^1 dx/x
[ \kappa_R(f_{u/\gamma}(y,M^2)) + f_{d/\gamma}(y,M^2)) +
\kappa_L f_{u/\gamma}(y,M^2) ]
f_{\gamma/e}(x)}
\end{eqnarray}
\end{itemize}
In all cases above, the momentum fraction $y$ of the quark within the photon
is given by $y = M^2/(x s)$ and $f_{\gamma/e}(x)$ is the backscattered laser
photon spectrum.  The negative sign in front of the vector LQ asymmetries is a
standard result: it comes about from the annihilation of 2 objects with
opposite helicity into a vector particle.  Another comment should be made at
this time.  Up to now, we've assumed that the beams will be polarized
perfectly.  This is probably not a bad assumption for the photon beam, as the
backscattered photon beam will carry the polarization of the incident laser
beam, and it should be straightforward to polarize the laser to a very high
degree.  Electron beam polarizations of order 70\% can be expected, and this
will modify some of our arguments.  First, even if the LQ couples only to a
particular helicity of electron, the contamination of the $e$ beam with the
wrong helicity will contaminate the signal.  The finite polarization of the
electron beam ($\lambda_{beam}$) will dilute the observable asymmetries by a
factor $\lambda_{beam}$.  Some care will have to be taken to ensure that any
LQ signal observed with polarized beams is real, that is {\bf not} due to
contamination of the beam.

Having estimates of event numbers from previous works, we now need to
determine if the different asymmetries predicted can be statistically
separated.  Due to a complete lack of data on parton distribution functions
within a polarized photon, we need to use some theoretical input on the shapes
of the helicity difference distribution functions of partons within the
photon.  There exist some parameterizations of the {\it asymptotic} polarized
photon distribution functions \cite{hassan,xu}, where it is assumed that $Q^2$
and $x$ are large enough that the Vector Meson Dominance part of the photon
structure is not important, but rather the behavior is dominated by the
point-like $\gamma q \bar{q}$ coupling.  In this sort of approximation, the
distribution functions take the form:
\begin{equation}
\Delta f_{q/\gamma}(x,Q^2) = \frac{\alpha}{\pi}
\ln \left( \frac{Q^2}{\Lambda^2} \right) \frac{1}{x} \Delta p(x)
\end{equation}
where $\Delta p(x)$ is a polynomial.  In order to be consistent, we will use a
similar asymptotic parameterization for the unpolarized photon distribution
functions as well \cite{nic}, even though various sets of more correct photon
distribution functions exist ({\it e.g.}, \cite{DO,DG,GRV,LAC}).  We will only
use this asymptotic approximation in the unpolarized case only for the
calculation of the asymmetry, where it is hoped that in taking a ratio of the
asymptotic polarized to the asymptotic unpolarized photon distribution
functions, the error introduced will be minimized; still, we suggest that our
results be considered cautiously at least in the relatively small LQ mass
region.  We note that in the asymptotic approximation, the unpolarized photon
distribution functions have (not unexpectedly) a similar form to the polarized
photon distribution functions:
\begin{equation}
f_{q/\gamma}(x,Q^2) = \frac{\alpha}{\pi}
\ln \left( \frac{Q^2}{\Lambda^2} \right) \frac{1}{x^{1.6}} p(x)
\end{equation}
where $p(x)$ is another polynomial.  Finally, in regards to the question of
the polarized photon distribution functions, a more careful calculation of the
helicity difference distributions does exist \cite{gv} which includes the
effect of VMD.  However the authors of Ref.~\cite{gv} give parameterizations
of the ratio of the helicity difference to the helicity sum distribution
functions that are independent of $Q^2$.  These are reported to be valid for
$10 \leq Q^2 \leq 100$~GeV$^2$.  Given the large mass of the LQs being
considered, the $Q^2$ is much too high to use these parameterizations.

We find that the asymmetry $A_{LL}$ depends only on the dimensionless variable
$M/\sqrt{s}$, though the event numbers depend on $M$ and $\sqrt{s}$
separately.  For all the following figures, a) corresponds to the results at a
500~GeV $e^+ e^-$ collider operating in $e \gamma$ mode and b) corresponds to
a 1~TeV $e^+ e^-$ collider operating in $e \gamma$ mode.  Throughout, we use
an integrated luminosity of 50~fb$^{-1}$/yr and the unpolarized photon
distributions of Gl\"uck, Reya and Vogt \cite{GRV} to estimate the number of
LQ events in a given LQ model.  Also, unless noted otherwise, our results are
for $\kappa_L = \kappa_R = 1$.  The first step in determining the leptoquark
couplings would be to use electron polarization to divide the leptoquarks into
subsets that couple only to left handed electrons, right handed electrons, or
to both.  Once this is done the asymmetry can be used to distinguish between
leptoquarks within these subsets.  We show, in Figure~1, $A_{LL}$ {\it vs.}
$M$ for the set of LQs which couple only to left-handed electrons, that is
$S_3$, $U_3$, $\tilde{V}_2$ and $\tilde{R}_2$ in the notation of BRW.  The
asymmetries for the vector LQs ($U_3$ and $\tilde{V}_2$) have been multiplied
by $-1$ in order to reduce the scale to the point that the structure in the
asymmetries is visible.  That is, the scalar LQ's have positive values for
$A_{LL}$ while the vector LQ's have negative values for $A_{LL}$.  In
Figure~2, we show $A_{LL}$ {\it vs.} $M$ for the set of LQs which couple only
to right-handed electrons, that is $\tilde{S}_1$ and $\tilde{U}_1$.  We again
multiply the vector LQ ($\tilde{U}_1$) asymmetry by $-1$.  Finally, in
Figures~3, 4 and 5 we show $A_{LL}$ {\it vs.} $M$ for the set of LQs which
couple to both helicities of electrons, that is $S_1$, $V_2$, $U_1$ and
$R_2$.  We again multiply the asymmetries of the vector LQs ($V_2$ and $U_1$)
by $-1$.  As the asymmetries for this set of LQs depend on the arbitrary
couplings $\kappa_L$ and $\kappa_R$, we show results for various values of the
$\kappa$s: in Figure~3, $\kappa_L = \kappa_R = 1$, in Figure~4,
$\kappa_L = 1/2$ and $\kappa_R = 1$ and in Figure~5, $\kappa_L = 1$ and
$\kappa_R = 1/2$.

Using earlier results, ({\it e.g.} Figure~3 of Ref.~\cite{mad}), event numbers
can be estimated and a statistical uncertainty in the measurement of $A_{LL}$
($\delta A_{LL}$) can also be estimated.  For an asymmetry
\begin{equation}
A = \frac{\sigma(\alpha) - \sigma(\beta)}{\sigma(\alpha) + \sigma(\beta)}
\end{equation}
the statistical uncertainty, $\delta A_{LL}$ is given by the expression
\begin{equation}
\delta A_{LL} = \frac{1 - A^2}{\sqrt{1 - A}}
\frac{1}{\sqrt{2 N(\alpha)}} = \sqrt{\frac{1 - A^2}{N_{TOT}}}
\end{equation}
where $N(\alpha)$ and $N_{TOT}$ are $L \sigma(\alpha)$ and $L (\sigma(\alpha) +
\sigma(\beta))$ respectively, with $L$ being the integrated luminosity.  These
estimated $\delta A_{LL}$ are shown in the error bars on Figures~1---5.  It
can be seen that it is quite easy to distinguish a vector LQ from a scalar LQ,
as $A_{LL}$ is large and positive for all the scalar LQs while it is large and
negative for vector LQs.  For a LQ which couples only to left-handed electrons
(see Figure~1), it will be possible to differentiate between the two possible
vector or scalar LQs essentially up to the kinematical limit (remembering that
the backscattered laser photon spectrum cuts off at an $x$ of about 0.83; thus
the maximum energy of an $e \gamma$ collider is slightly lower than the energy
of the corresponding $e^+ e^-$ collider).  If the LQ couples only to
right-handed electrons (see Figure~2), there is only one vector and one scalar
LQ possible, so it will be possible to determine the particular LQ essentially
up to the kinematical limit.  Finally, for a LQ which couples to both helicity
electrons (see Figures~3, 4 and 5), it will be possible to differentiate
between the two possible vector or scalar LQs essentially up to the
kinematical limit, except for a small region of LQ mass where the asymmetries
for the two types of scalar LQ are equal.  The precise mass at which the
crossover occurs depends on the values of $\kappa_{L,R}$.  The limits quoted
here assume a fully polarized electron beam; finite polarization will reduce
these limits somewhat.

In conclusion, it certainly appears that a polarized $e \gamma$ collider can
be used to differentiate between the different models of LQs that can exist,
essentially up to the kinematic limit of the $e \gamma$ collider.
Furthermore, it is quite straightforward to distinguish scalar LQs from vector
LQs for all LQ mass (given that the LQ is kinematically allowed) with only a
polarized electron beam.  It is thus clear that a more careful analysis is
warranted, with the following improvements:
\begin{itemize}
\item{We rely on theoretical input for information on the parton distribution
functions within a polarized photon; given the EMC (proton) spin crisis, there
may be some surprises in the polarized photon as well.}
\item{There are many questions as to the reliability of the asymptotic
approximation to the photon distribution functions: are the values of $y$ (the
momentumn fraction of the quark within the photon) and $Q^2$ large enough that
the photon behaves asymptotically?  Is it possible, at the very least (in the
absence of experimental data) to improve the theoretical input into the
polarized photon distribution functions?}
\end{itemize}

\acknowledgments

This research was supported in part by the Natural Sciences and Engineering
Research Council of Canada. The authors are grateful to Manuel Drees and Drew
Peterson for helpful communications, to JoAnne Hewett for suggesting this
particular analysis and to Tom Rizzo for continual encouragement.

\figure{$A_{LL}$ {\it vs.} $M$ for LQs which couple only to left-handed
electrons; a) is for a 500~GeV collider and b) is for a 1~TeV collider.  The
solid curve is for an $S_3$ LQ, the dashed line is for a $\tilde{V}_2$ LQ
($-A_{LL}$ shown), the dotted line is for a $U_3$ LQ ($-A_{LL}$ shown) and the
dotdashed line is for a $\tilde{R}_2$ LQ.}

\figure{$A_{LL}$ {\it vs.} $M$ for LQs which couple only to right-handed
electrons; a) is for a 500~GeV collider and b) is for a 1~TeV collider.  The
solid curve is for an $\tilde{S}_1$ LQ and the dashed line is for a
$\tilde{U}_1$ LQ ($-A_{LL}$ shown).}

\figure{$A_{LL}$ {\it vs.} $M$ for LQs which couple only to both left- and
right-handed electrons; a) is for a 500~GeV collider and b) is for a 1~TeV
collider; here $\kappa_L = \kappa_R = 1$.  The solid curve is for an $S_1$ LQ,
the dashed line is for a $V_2$ LQ ($-A_{LL}$ shown), the dotted line is
for a $U_1$ LQ ($-A_{LL}$ shown) and the dotdashed line is for a $R_2$ LQ.}

\figure{$A_{LL}$ {\it vs.} $M$ for LQs which couple only to both left- and
right-handed electrons; a) is for a 500~GeV collider and b) is for a 1~TeV
collider; here $\kappa_L = 0.5$ and $\kappa_R = 1$.  The solid curve is for an
$S_1$ LQ, the dashed line is for a $V_2$ LQ ($-A_{LL}$ shown), the dotted line
is for a $U_1$ LQ ($-A_{LL}$ shown) and the dotdashed line is for a $R_2$ LQ.}

\figure{$A_{LL}$ {\it vs.} $M$ for LQs which couple only to both left- and
right-handed electrons; a) is for a 500~GeV collider and b) is for a 1~TeV
collider; here $\kappa_L = 1$ and $\kappa_R = 0.5$.  The solid curve is for an
$S_1$ LQ, the dashed line is for a $V_2$ LQ ($-A_{LL}$ shown), the dotted line
is for a $U_1$ LQ ($-A_{LL}$ shown) and the dotdashed line is for a $R_2$ LQ.}

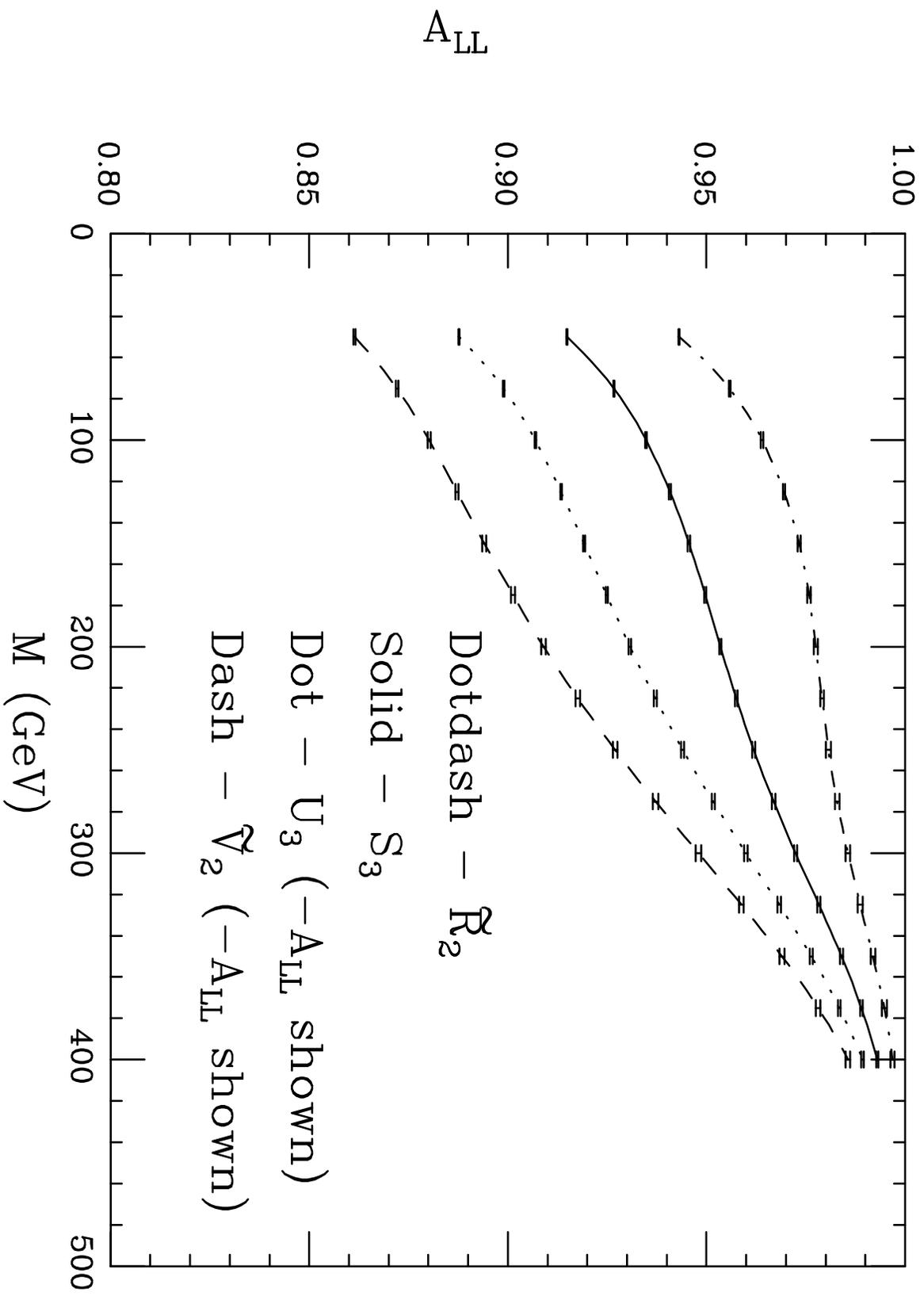
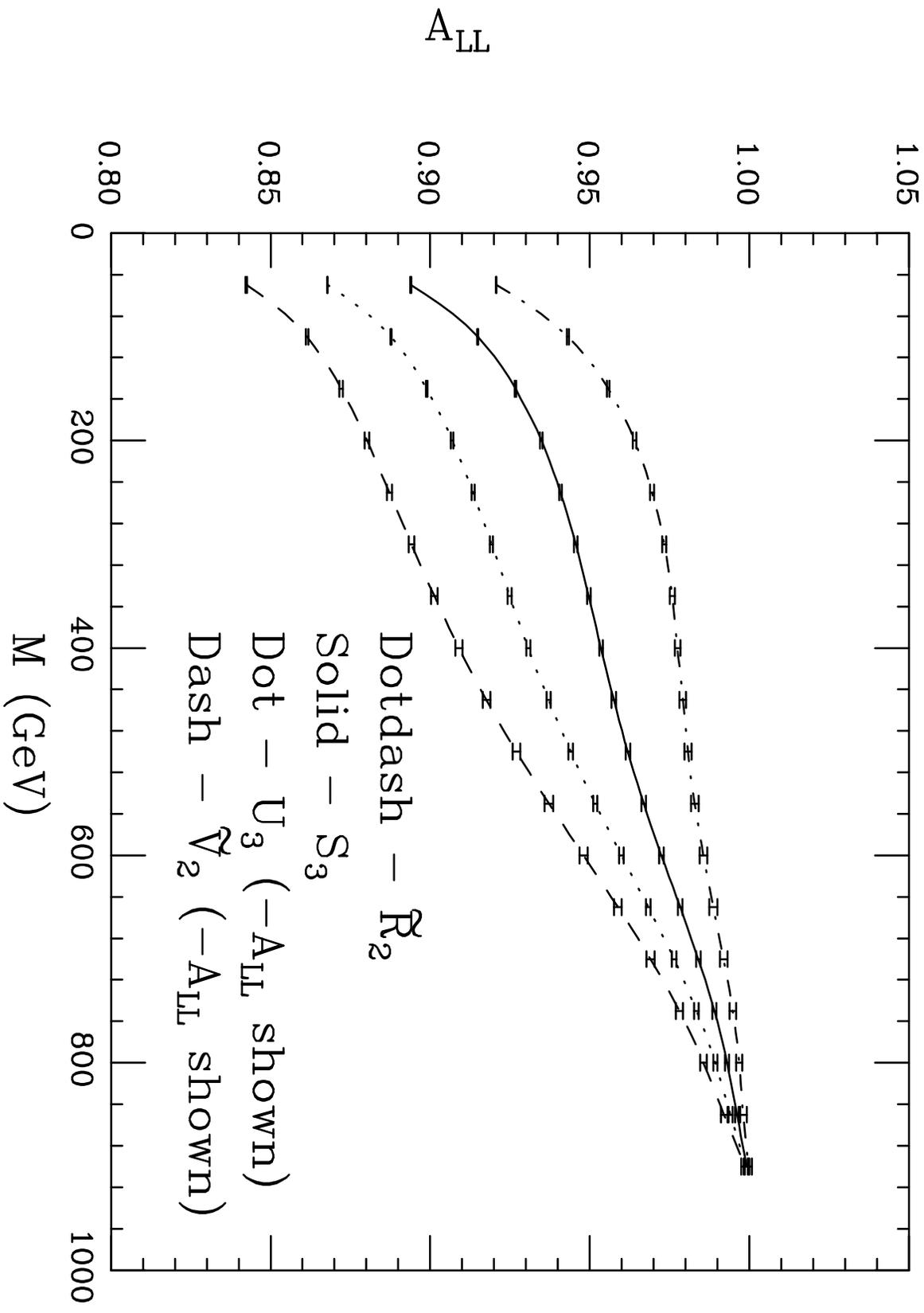
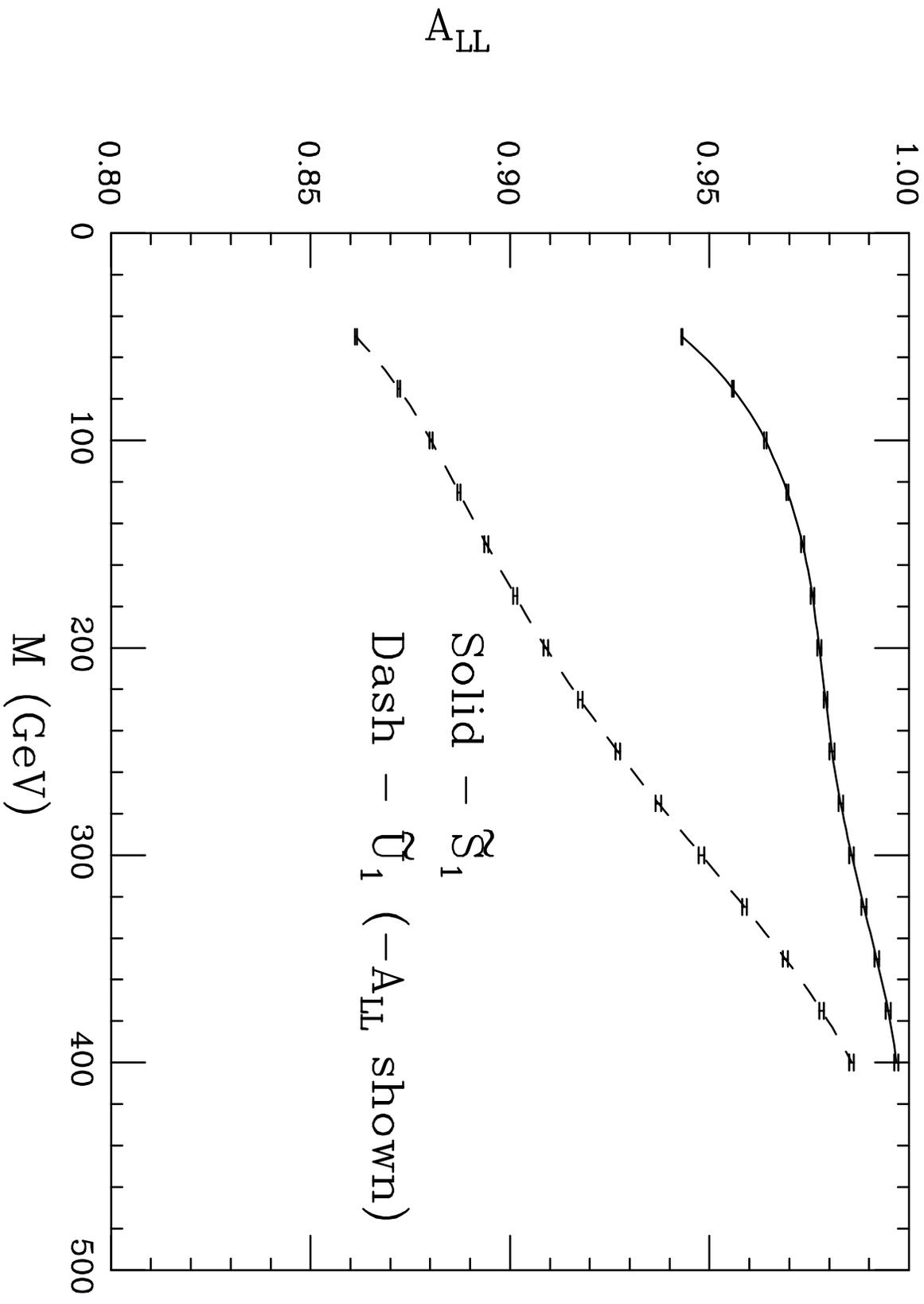
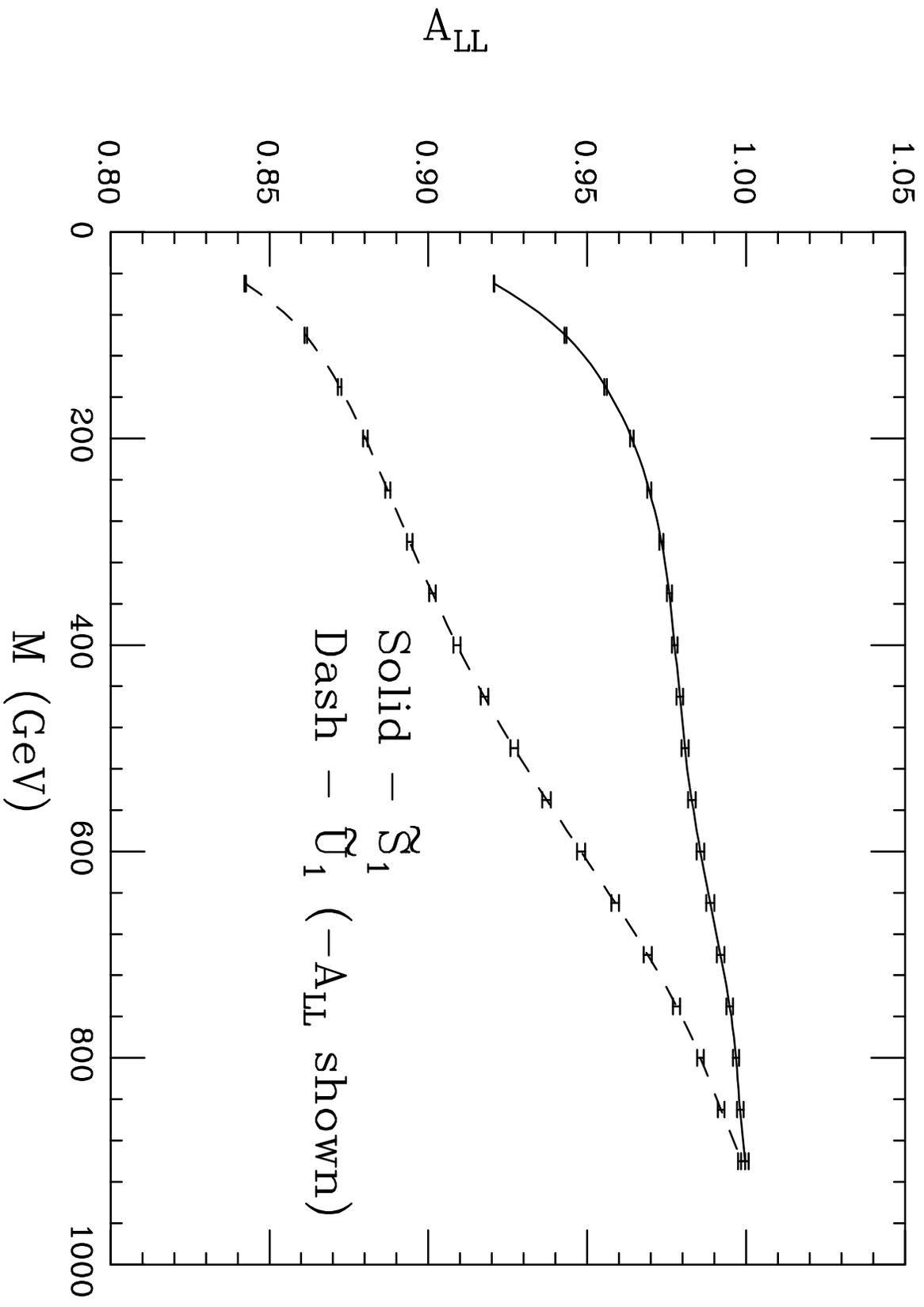
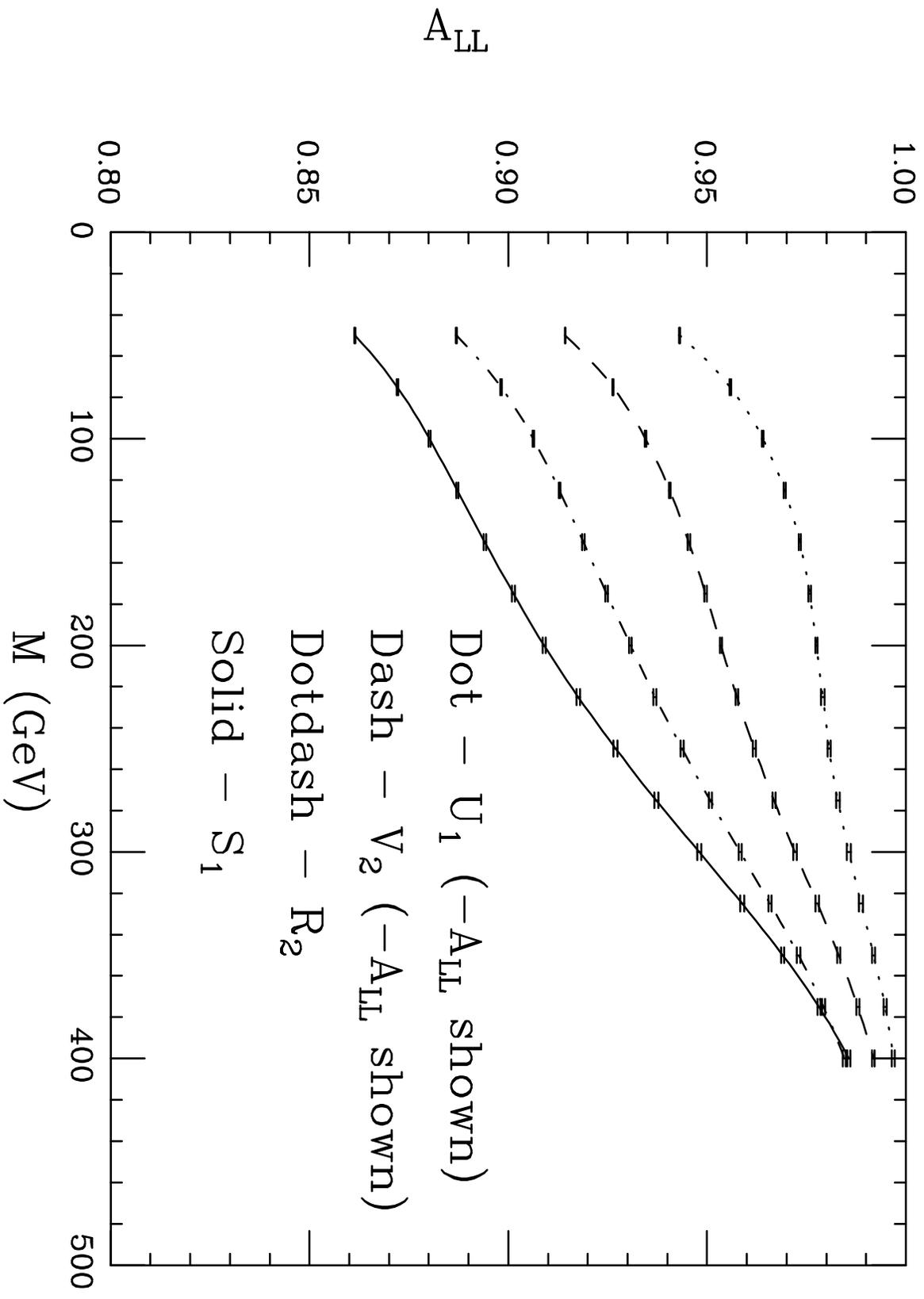
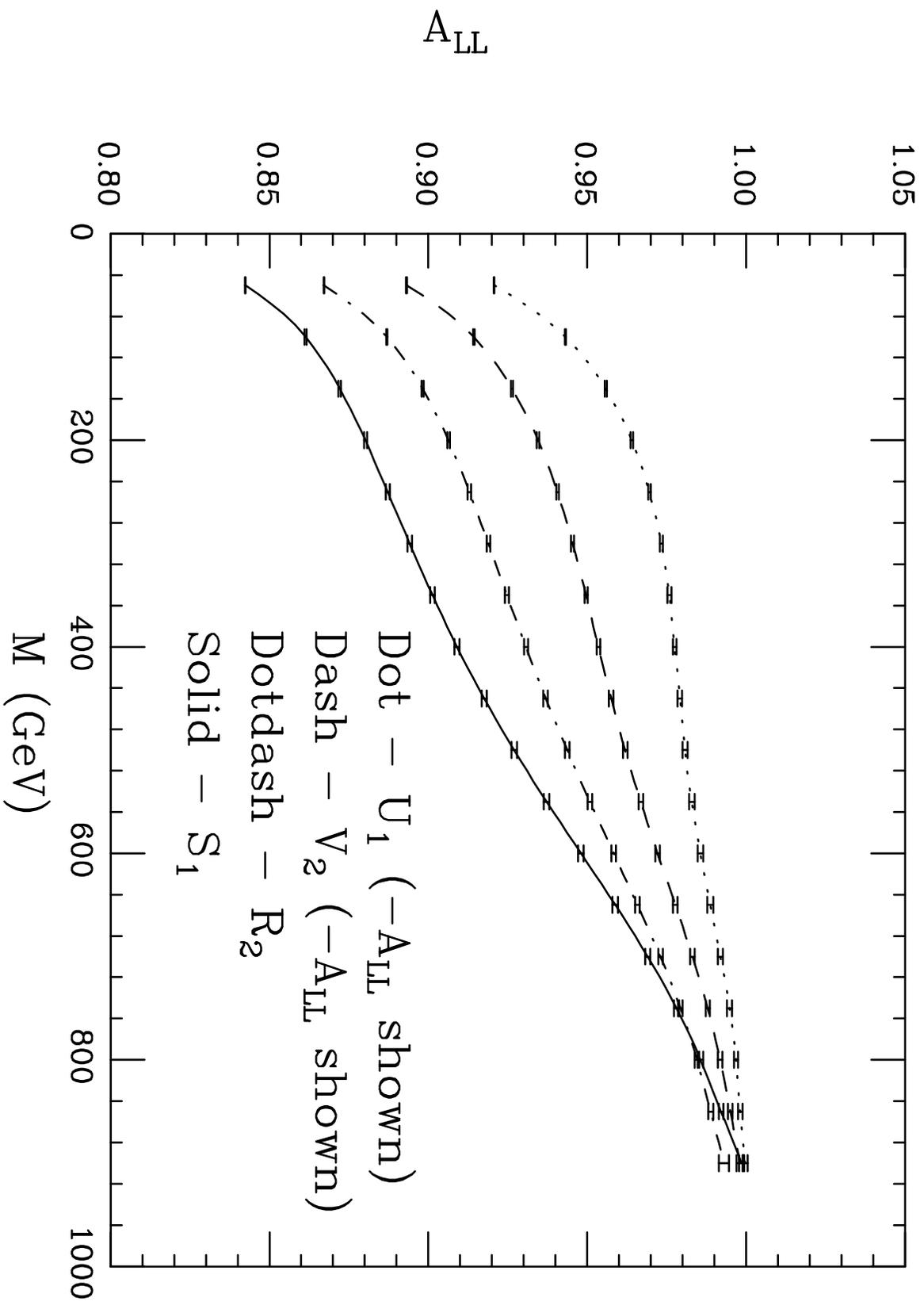
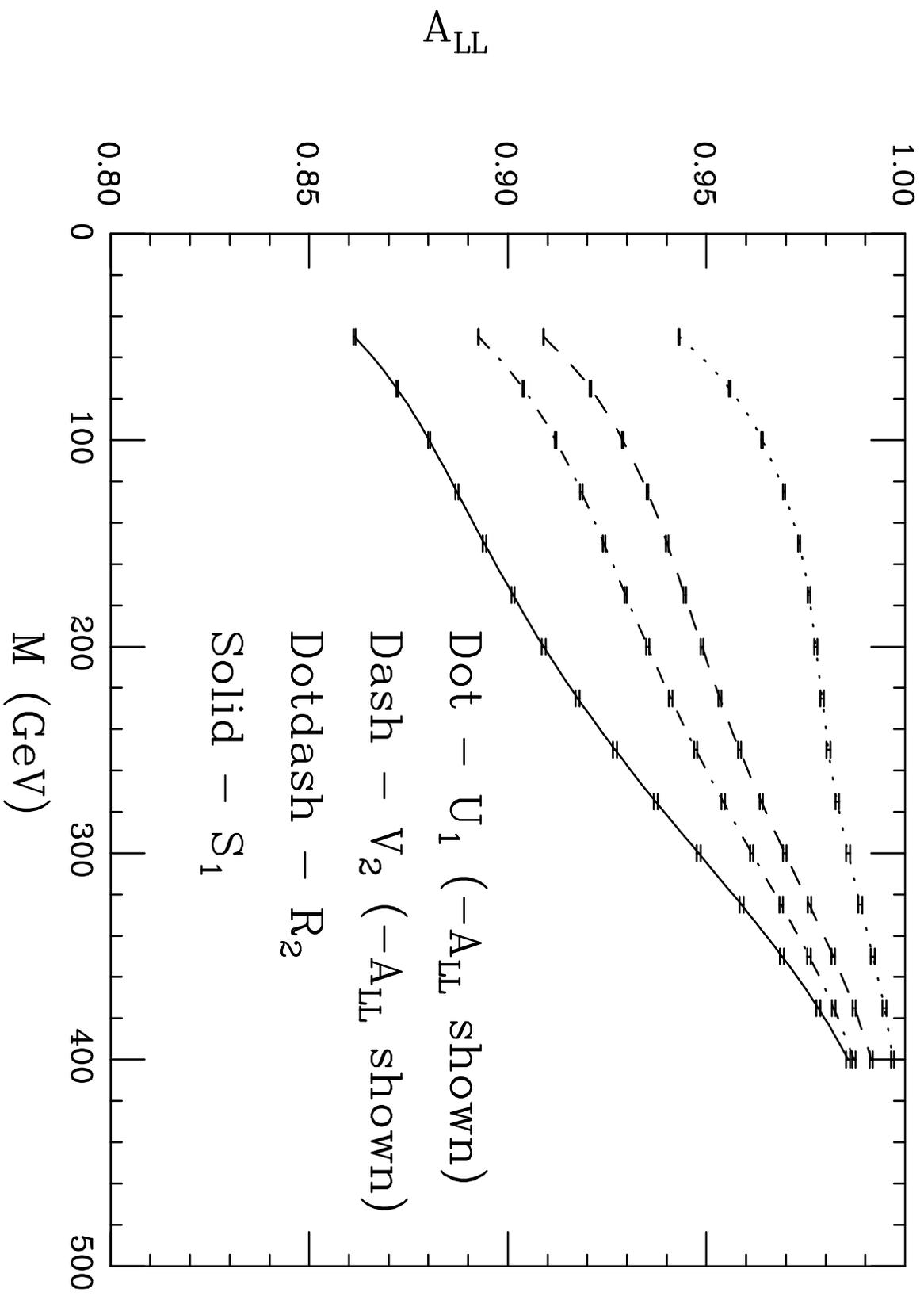
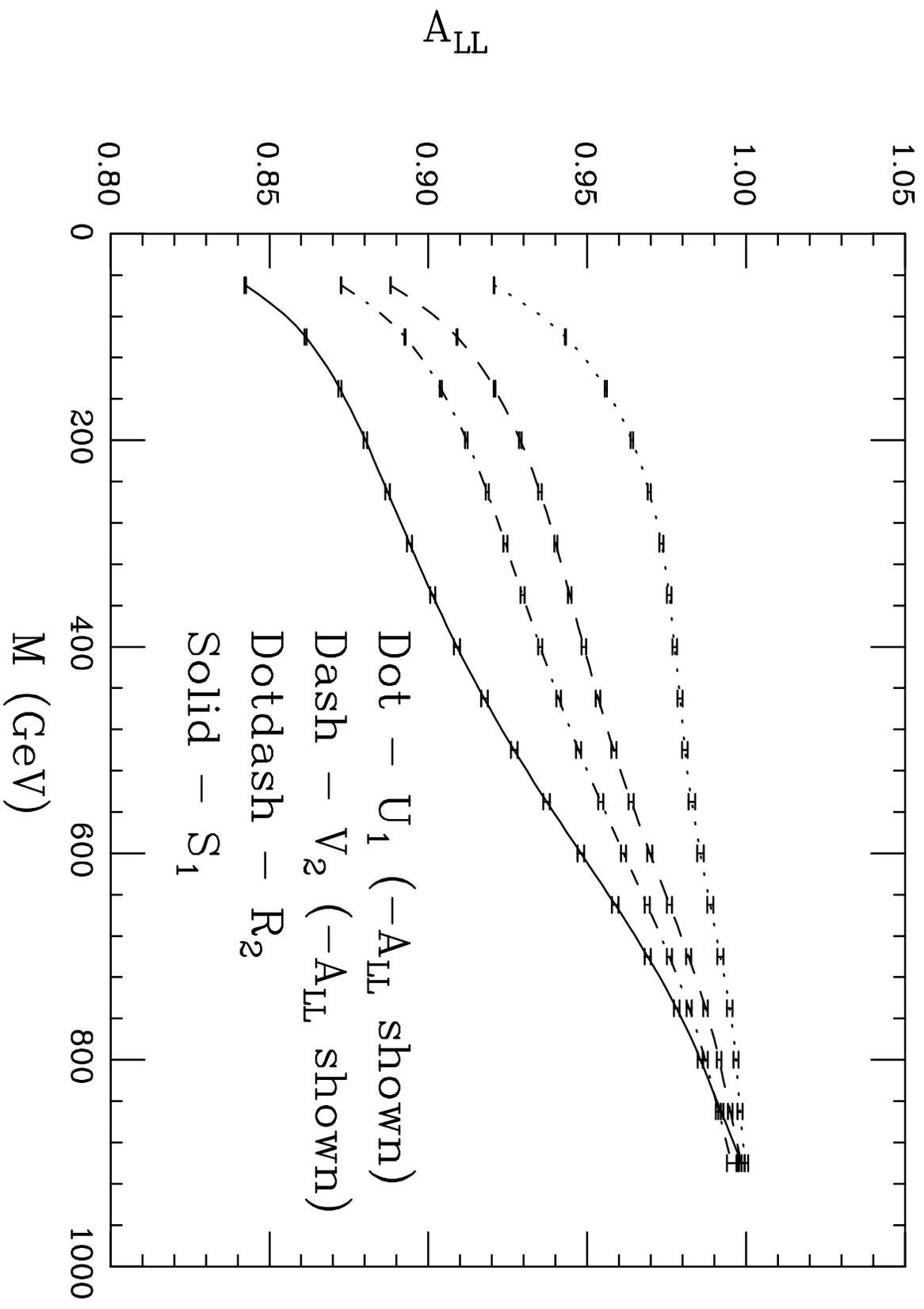
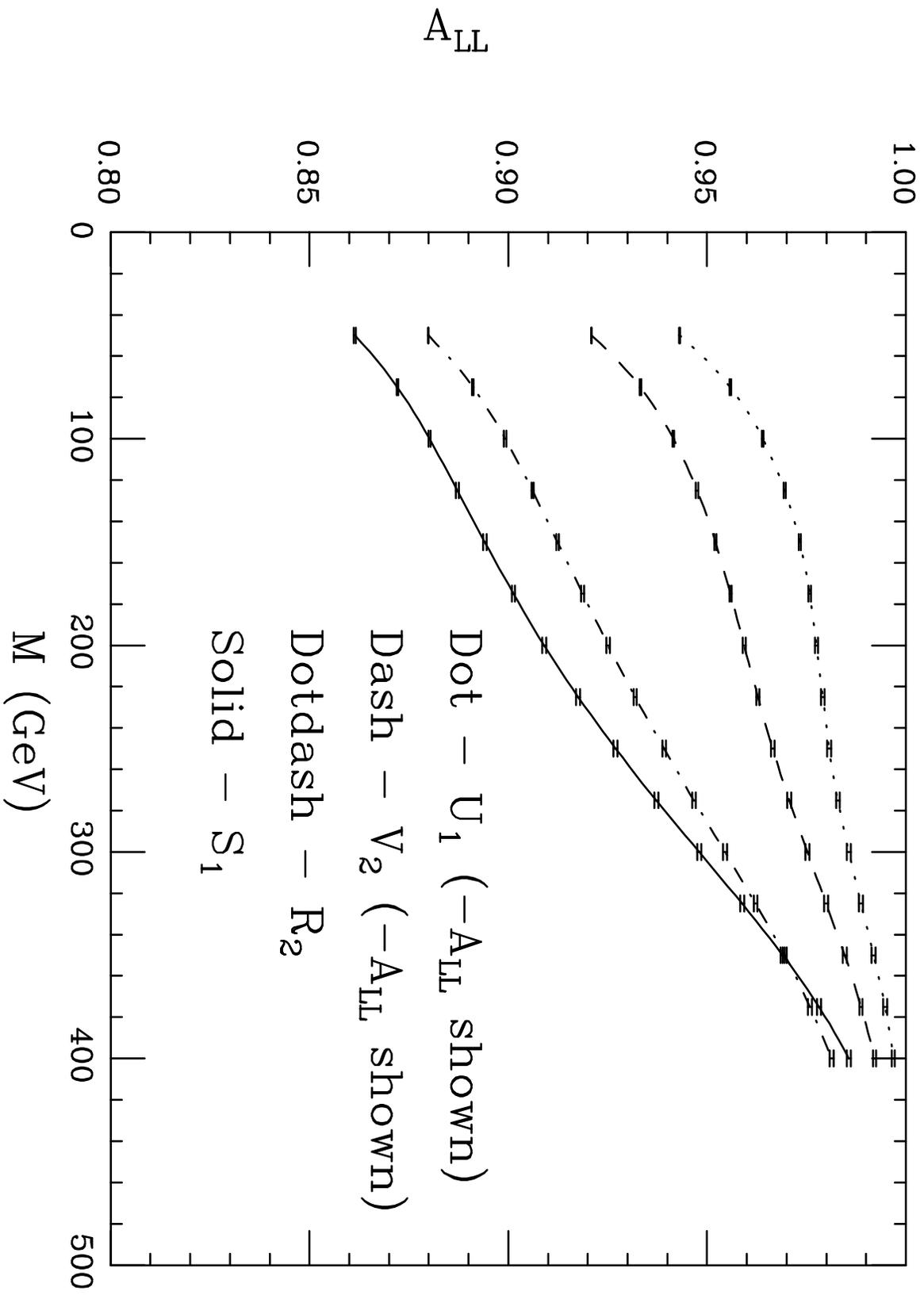
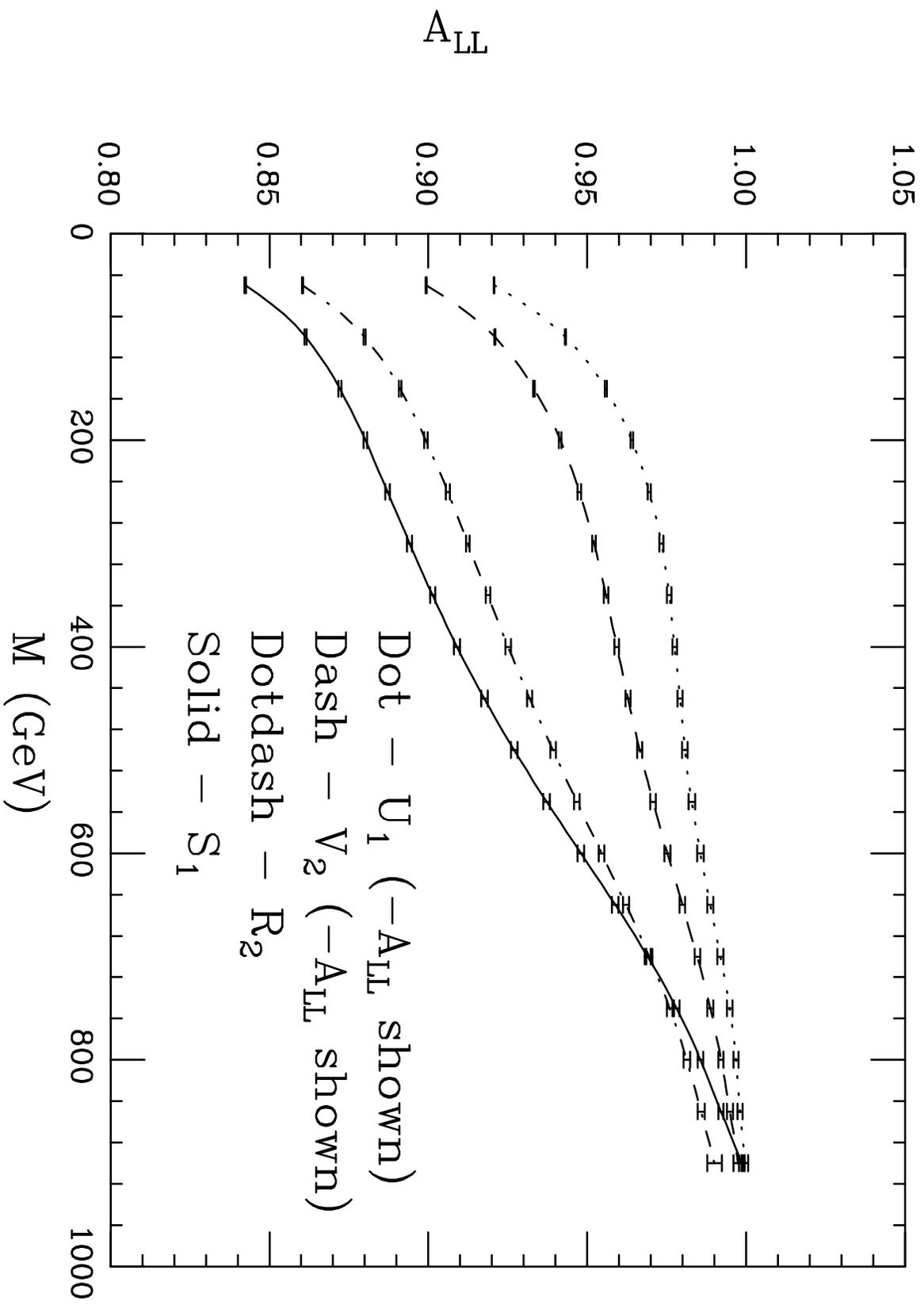
\end{document}